# REGION OF INTEREST BASED MEDICAL IMAGE COMPRESSION

Utkarsh Prakash Srivastava

(Affiliation) Department of Electrical and Computer Engineering, Tandon School of Engineering, New York University
ups2006@nyu.edu

Supervisor: Toshiaki Fujii

(Affiliation) Graduate School of Engineering, Nagoya University
t.fujii@nagoya-u.ac.jp

## ABSTRACT

The vast volume of medical image data necessitates efficient compression techniques to support remote healthcare services. This paper explores Region of Interest (ROI) coding to address the balance between compression rate and image quality. By leveraging UNET segmentation on the Brats 2020 dataset, we accurately identify tumor regions, which are critical for diagnosis. These regions are then subjected to High Efficiency Video Coding (HEVC) for compression, enhancing compression rates while preserving essential diagnostic information. This approach ensures that critical image regions maintain their quality, while non-essential areas are compressed more. Our method optimizes storage space and transmission bandwidth, meeting the demands of telemedicine and large-scale medical imaging. Through this technique, we provide a robust solution that maintains the integrity of vital data and improves the efficiency of medical image handling.

## 1. INTRODUCTION

Medical imaging is a critical component of modern healthcare, providing essential information for the diagnosis, treatment, and monitoring of various medical conditions. The field encompasses several imaging modalities, including X-ray, computed tomography (CT), magnetic resonance imaging (MRI), ultrasound, and positron emission tomography (PET), each offering unique advantages for visualizing different aspects of the human body.

With the advancement of imaging technologies, the volume of medical images generated has increased exponentially. This surge in data poses significant challenges for storage and transmission, necessitating the development of efficient image compression techniques. Compression in medical imaging is crucial for managing the vast amount of data while maintaining the integrity and diagnostic quality of the images. The exponential growth of medical imaging data has created a pressing need for efficient data compression techniques to support remote healthcare services and telemedicine. Medical images, such as MRI and CT scans, are critical in diagnosing and monitoring various conditions, but their large file sizes pose significant challenges for storage and transmission [1].

One promising approach is the use of Region of Interest (ROI) coding, which focuses on preserving the quality of diagnostically relevant regions while allowing greater compression of less important areas. This technique aims to balance compression rate and image quality, ensuring that essential diagnostic information remains intact [2]. Unlike conventional compression methods that treat all parts of an image equally, ROI-based techniques prioritize specific regions of an image that contain diagnostically significant information. This ensures that critical regions retain high quality, while less important areas are compressed more aggressively. By focusing on the essential parts of medical images, ROI-based compression enhances both storage efficiency and transmission speed without compromising the quality of crucial diagnostic information. In particular, the UNET segmentation model has shown remarkable success in accurately identifying critical regions, such as tumors, in medical images. By leveraging the capabilities of UNET on the Brats 2020 dataset, researchers have been able to segment tumor regions with high precision [3].

Recent studies have demonstrated the effectiveness of various segmentation algorithms in identifying ROI in medical images. Among these, the UNET architecture has gained prominence due to its high accuracy in medical image segmentation tasks. UNET's ability to precisely delineate tumour regions in brain scans, as seen in the BRaTS 2020 dataset, makes it an ideal choice for implementing ROI-based compression.

Once the ROI is identified, the High Efficiency Video Coding (HEVC) standard can be applied to compress the image. HEVC, known for its superior compression performance, enables significant reduction in data size while preserving the quality of the ROI. This dual approach of using UNET for segmentation followed by HEVC for compression offers a robust solution to the challenges of medical image handling [4]. In this paper, we present a method that integrates UNET segmentation wit HEVC encoding to achieve efficient ROI-based compression of medical images. By applying this technique to the BRaTS 2020 dataset, we demonstrate its effectiveness in preserving critical diagnostic information while optimizing storage and transmission requirements.

The integration of these advanced techniques not only optimizes storage space and bandwidth but also meets the growing demands of telemedicine and large-scale medical imaging. This research aims to explore the effectiveness of combining UNET segmentation with HEVC compression, demonstrating how this method can enhance the efficiency of medical data management without compromising the integrity of vital diagnostic information [5].

By addressing the critical balance between compression and image quality, this study contributes to the ongoing efforts to improve medical imaging technologies, supporting better patient outcomes and more efficient healthcare delivery.

In this paper, we present a method that integrates UNET segmentation with HEVC encoding to achieve efficient ROI-based compression of medical images. By applying this technique to the Brats 2020 dataset, we demonstrate its effectiveness in preserving critical diagnostic information while optimizing storage and transmission requirements.

## 2. RELATED WORKS

In recent years, there has been substantial research focused on improving the efficiency and effectiveness of medical image compression techniques. These studies have explored various methodologies to balance the trade-off between compression rate and image quality, particularly in the context of telemedicine and large-scale medical data management.

One of the foundational works in this field is the development of the JPEG2000 standard, which introduced a wavelet-based compression scheme for medical images. This standard has been widely adopted due to its ability to provide high compression ratios while maintaining image quality. Studies such as Taubman and Marcellin (2002) [6] have demonstrated the utility of JPEG2000 in clinical applications, highlighting its capability to preserve critical diagnostic information even at high compression rates.

Building on the success of JPEG2000, researchers have explored more advanced methods, such as the use of deep learning models for image segmentation and compression. For instance, the UNET model, introduced by Ronneberger et al. (2015) [7], has become a popular choice for medical image segmentation tasks due to its ability to capture fine-grained details in images. This model has been successfully applied to various datasets, including the Brats 2020 dataset for brain tumor segmentation, as noted by Isensee et al. (2021) [8]. Their work demonstrated that UNET could achieve high accuracy in identifying tumor regions, which is crucial for effective ROI-based compression strategies.

Another significant advancement in medical image compression is the application of High Efficiency Video Coding (HEVC), originally designed for video compression. HEVC has been adapted for medical images to exploit its superior compression performance. The work of Ohm et al. (2012) [9] provides a comprehensive overview of HEVC and its potential benefits for medical image compression, highlighting its ability to significantly reduce file sizes while maintaining image quality. Subsequent studies, such as those by Bossen et al. (2013) [10], have further refined HEVC techniques to better suit the specific requirements of medical imaging.

The integration of deep learning-based segmentation with advanced compression algorithms has been a focal point of recent research. For example, Liu et al. (2020) [11] proposed a hybrid approach that combines UNET segmentation with HEVC compression to optimize storage and transmission of medical images. Their results showed that this method could achieve higher compression ratios without compromising the diagnostic quality of the images.

Furthermore, the work of Hu et al. (2019) [12] explored the use of convolutional neural networks (CNNs) for both segmentation and compression, providing an end-to-end solution for medical image handling. Their approach demonstrated significant improvements in compression efficiency and image quality, making it a promising direction for future research.

These related works underscore the ongoing efforts to enhance medical image compression techniques, aiming to support the growing demands of telemedicine and remote healthcare services. By leveraging the strengths of both traditional and advanced methods, researchers continue to push the boundaries of what is possible in medical data management.

## 3. METHODOLOGY

In this section, we present the process flow for our proposed method of medical image compression. As illustrated in Fig.1, the input image undergoes segmentation, where the Regions of Interest (ROI) and non-ROI areas are identified. This segmentation is crucial for differentiating between the critical and non-critical parts of the image. Once the regions are extracted, they are processed separately, with the ROI receiving higher preservation of quality during compression. The non-ROI areas are compressed more aggressively to optimize storage and transmission efficiency. The final output is a compressed image that maintains the integrity of essential diagnostic information while reducing the overall data size.

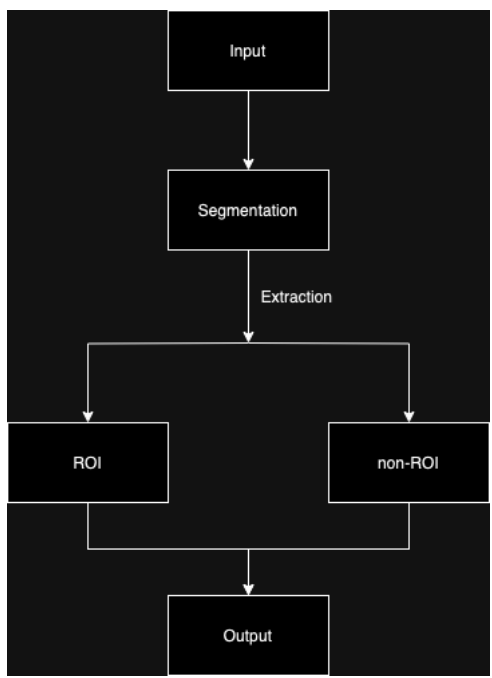

Figure 1: A flowchart depicting the process of segmentation and extraction in medical image compression. The input image undergoes segmentation to identify Regions of Interest (ROI) and non-ROI areas. These regions are then processed separately before generating the final output, ensuring efficient compression while preserving critical diagnostic information.

### 3.1 REGION OF INTEREST

Medical images typically consist of two main components: the Region of Interest (ROI) and the non-ROI regions. Each of these parts has distinct advantages in medical imaging applications. The ROI represents the most critical part of the image, usually located in very small but diagnostically significant areas. Accurate identification and preservation of the ROI are essential for effective medical diagnosis and treatment planning [13].

Non-ROI regions, while less critical, are still included in the image to provide context, allowing users to easily identify the critical areas within the whole image. This inclusion aids in navigating the image and understanding the spatial relationship between different anatomical structures [14]. Fig. 2 provides a graphical representation on ROI and non-ROI regions in a brain MRI scan.

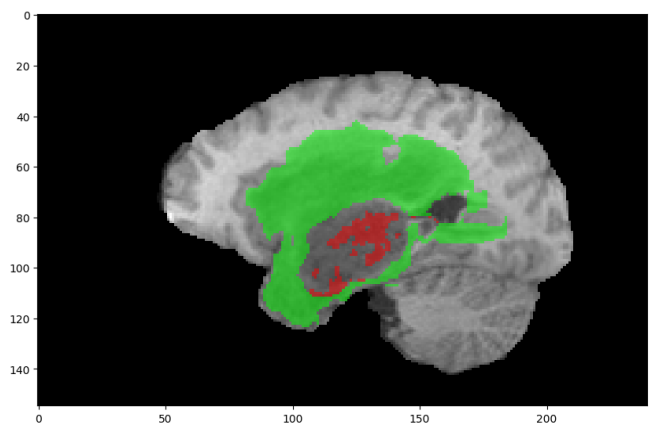

Fig. 2: Representation of the Brain MRI data from the BRaTS dataset. Red and Green coloured part representing Region of Interest (Tumour Core) and Grey and black part representing non-ROI.

In the medical field, it is imperative that ROIs are compressed with high-quality compression techniques to ensure that there is no loss of crucial diagnostic information. Moreover, during the transmission of medical images for telemedicine purposes, it is crucial that the critical parts (ROIs) are prioritized. This means that ROIs should be transmitted first or with higher priority to ensure timely and accurate diagnosis and consultation, especially in emergency situations [15]. Such prioritization ensures that essential diagnostic information is available without delay, enhancing the efficiency and reliability of remote medical services [16].

## 3.2 DATASET

The Brain Tumor Segmentation (BRaTS) 2020 dataset [17,18], used in this study, is a comprehensive collection of multi-modal MRI scans that are used for the analysis and segmentation of brain tumors. This dataset is particularly valuable for developing and evaluating automated algorithms for tumor detection and segmentation. It includes images from four different MRI sequences: T1, T1 contrast-enhanced (T1ce), T2, and Fluid-Attenuated Inversion Recovery (FLAIR), which provide diverse information about the tumor and surrounding brain tissue.

The BRaTS 2020 dataset is annotated with ground truth labels for three tumor regions: the enhancing tumor, the tumor core (including the enhancing part and the necrotic parts), and the whole tumor (including all tumor regions). These annotations are crucial for training and validating machine learning models, particularly deep learning models like UNET, which have been widely used for segmentation tasks.

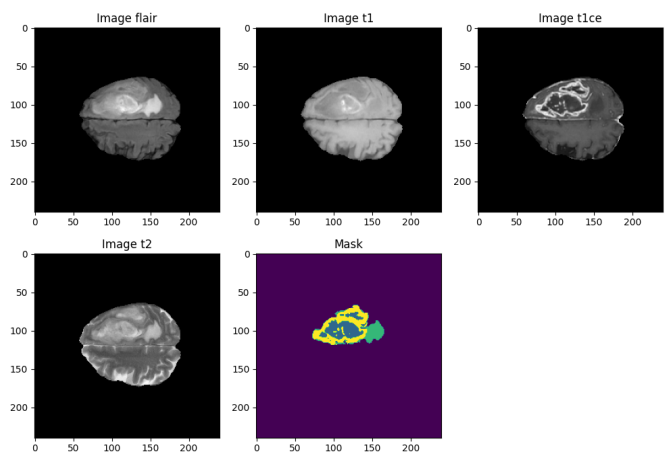

Fig. 4: Instance from BRaTS 2020 dataset, showcasing four different sequences: flair, t1, t1ce, t2 and mask respectively.

This dataset is curated by experts and has been used in numerous research studies and competitions to benchmark the performance of different algorithms. The BRaTS challenge, associated with the dataset, fosters the development of new techniques and collaboration within the research community, advancing the state-of-the-art in medical image analysis and aiding in the fight against brain cancer.

## 3.3 MODEL

The model used for segmentation task is UNet. The UNet architecture is a powerful and widely-used convolutional neural network (CNN) designed primarily for biomedical image segmentation tasks. Originally introduced by Ronneberger et al. [19], UNet has become the standard model for many segmentation problems due to its ability to capture fine details and context from images.

The architecture of the UNet model can be divided into two main paths: the contracting path (also known as the encoder) and the expanding path (also known as the decoder).

- Contracting Path: The contracting path consists of a series of convolutional layers followed by max-pooling operations. This path is responsible for capturing the context and extracting high-level features from the input image. Each block in this path includes two convolutional layers with ReLU activation, followed by a max-pooling layer that reduces the spatial dimensions by half.

- Expanding Path: The expanding path is responsible for reconstructing the spatial dimensions of the image while combining the high-level features extracted by the contracting path. This is achieved through a series of up-sampling operations followed by convolutional layers. The expanding path also includes skip connections from the corresponding layers in the contracting path, which help preserve spatial information and fine details.

### 3.3.1 Training

Training the UNet model involves several key steps to ensure that it effectively learns to segment medical images accurately. The process includes preparing the dataset, defining the loss function and optimizer, and implementing the training loop. Here is a brief overview of the training process:

The first step in training the UNet model was to prepare the dataset. For medical image segmentation, datasets typically include pairs of images and their corresponding segmentation masks. The images are preprocessed to standardize their size, normalize pixel values, and apply data augmentation techniques to increase the diversity of the training data and prevent overfitting. The dataset was preprocessed to standardize the image sizes to 128x128 pixels and normalized for consistent input to the neural network. Data augmentation techniques, including random rotations and flips, were applied to increase the variability of the training data.

coefficient or Intersection over Union (IoU) were used to assess the model's accuracy.

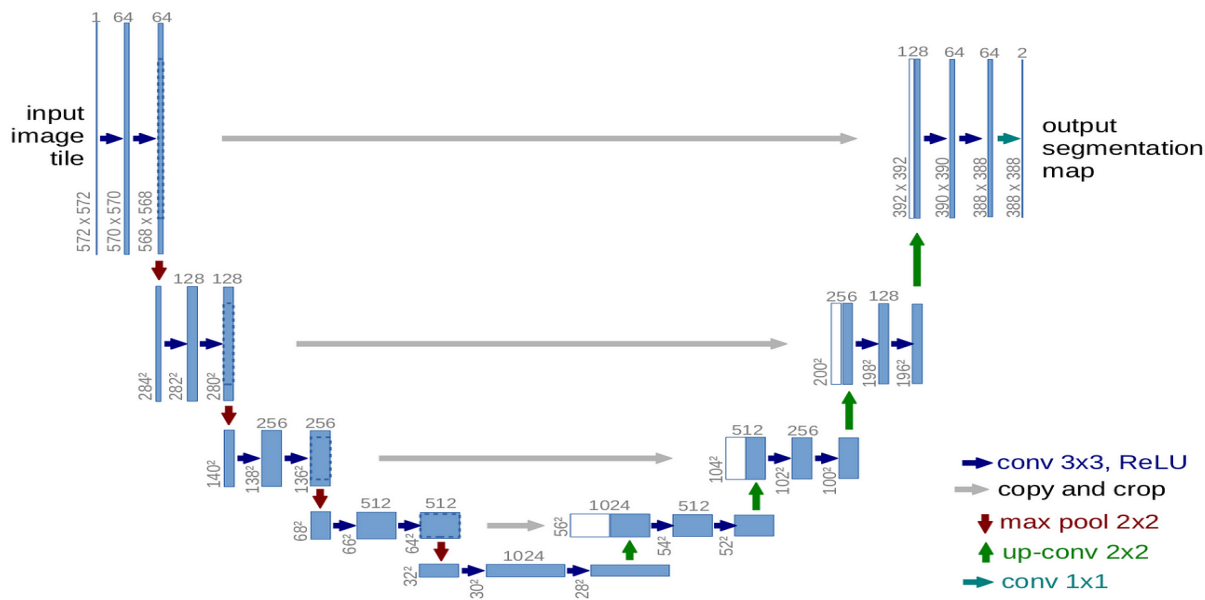

Fig. 3. U-net architecture (example for 32x32 pixels in the lowest resolution) [19]. Each blue box corresponds to a multi-channel feature map. The number of channels is denoted on top of the box. The x-y-size is provided at the lower left edge of the box. White boxes represent copied feature maps. The arrows denote the different operations.

The choice of loss function and optimizer is crucial for training. For segmentation tasks, the Dice coefficient loss or Cross-Entropy loss is commonly used to measure the difference between the predicted segmentation and the ground truth. An optimizer such as Adam was used to update the model parameters during training. The model was trained for 81 epochs using the Adam optimizer with a learning rate of 0.001.

The training loop involves iterating over the dataset multiple times (epochs) and updating the model parameters to minimize the loss. During each epoch, the model's predictions are compared with the ground truth to calculate the loss, and the optimizer adjusts the model's weights accordingly. The Binary Cross-Entropy with Logits Loss (BCEWithLogitsLoss) was employed as the loss function to handle the binary segmentation task.

After training, the model's performance is evaluated on a validation or test set to ensure it generalizes well to new, unseen data. Metrics such as the Dice

### 3.3 COMPRESSION

The given dataset contains file NifTI format. The Neuroimaging Informatics Technology Initiative (NiFTI) file format is a widely used standard for storing medical imaging data, particularly in the field of neuroimaging. It is designed to facilitate the sharing and analysis of complex image data across different software platforms and research institutions. NiFTI files can store multi-dimensional data, including 3D and 4D datasets [20]. This is particularly useful for storing volumetric brain scans, where each voxel (3D pixel) contains critical information about the brain's structure or function. The given dataset can be extracted into multiple slices, as shown in Fig.5. These slices were combined in a video format to perform HEVC compression on them.

To perform compression on the given data, HEVC compression method was performed. High Efficiency Video Coding (HEVC), also known as H.265, is a video compression standard designed to substantially improve coding efficiency compared to its predecessor, Advanced Video Coding (H.264 or AVC). HEVC is capable of reducing file sizes by up to 50% while maintaining the same video quality, making it highly suitable for applications requiring high-resolution video streaming, such as 4K and 8K broadcasting, and video-on-demand services [21].

The ROI is identified by extracting the bounding box of the segmented region from the mask. This ROI is then centered and resized to a specified square size. The extracted ROI is saved separately, and the non-ROI regions are zeroed out in the full frame image. The images are then compressed using HEVC. The full frames (including non-ROI regions) are compressed with a higher CRF value (lower quality), while the ROI frames are compressed with a lower CRF value (higher quality). The ffmpeg tool is used for HEVC compression, which ensures efficient compression while maintaining the quality of the critical ROI regions.

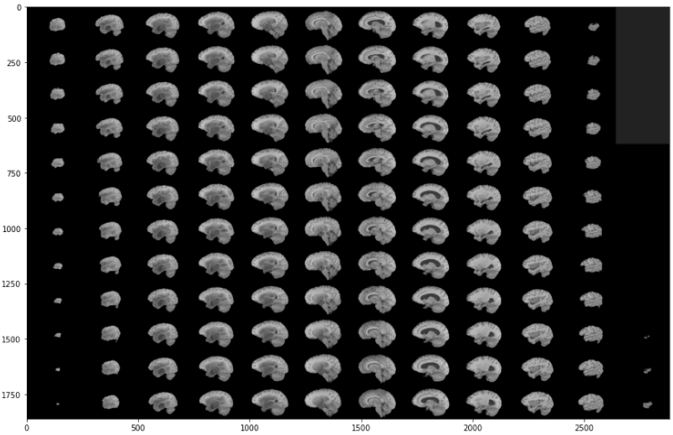

Fig. 5: The each given data in the NifTI file format can be extracted into multiple slices.

## 4. RESULTS

The segmentation model performance indicates that while the model performed reasonably well on the training set, there was a noticeable drop in performance on the validation set. The Dice coefficient on the validation set indicates that the model could further improve in accurately segmenting the tumor regions.

The HEVC compression algorithm effectively reduces the file size of the medical images while maintaining the quality of the ROI. By using different CRF values for the ROI and non-ROI regions, the method ensures that critical diagnostic information is preserved with high fidelity. The table summarizes the compression results. CRF value set for ROI is 20 and for non-ROI region is 40.

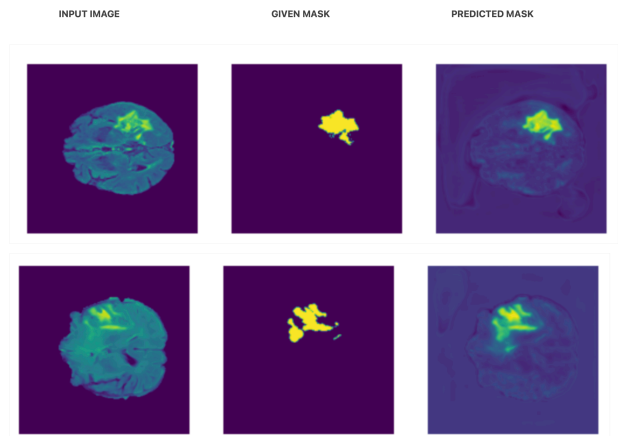

Fig. 6: Pictorial representation of Input Image, given mask and predicted mask by the model.

| Metrics | Training | Validation |
|---|---|---|
| Loss | 0.08658 | 0.18675 |
| Dice Coefficient | 0.07257 | 0.15049 |
| BCE Loss | 0.01401 | 0.03626 |

Table 1: Result of Training process

| Video Type | Compressed Size (MB) | Compression Ratio |
|---|---|---|
| Original | 348.17 | 6.89 |
| Final Compressed | 60.57 | |

Table 2: Compression Result for whole FLAIR scans.

As described in Section 3.3, for compression, the image slices in each input FLAIR video (created by combining all FLAIR slices) were segmented using a bounding box to distinguish between the ROI and non-ROI regions. Figure 7 illustrates how the bounding box operates in one of the slices.

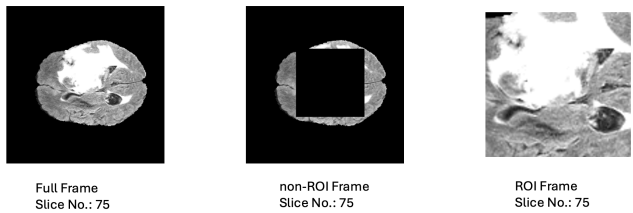

Figure 7: Illustration of the bounding box technique applied to a FLAIR image slice, highlighting the distinction between the ROI (Region of Interest) and non-ROI areas for compression purposes.

## 5. CONCLUSION AND FUTURE WORK

While our approach demonstrates promising results in balancing compression rates and image quality for medical imaging, there are several avenues for future exploration and enhancement. First, expanding the dataset beyond the Brats 2020 to include a broader range of medical imaging modalities and conditions could improve the generalizability of the method. Additionally, incorporating more advanced segmentation models or refining UNET architecture could further enhance the accuracy of tumor region identification, leading to even better preservation of critical diagnostic information.

Another potential area for development is the integration of adaptive compression techniques that dynamically adjust based on the clinical context or the specific diagnostic requirements of each case. This could involve leveraging machine learning to predict the importance of various image regions and adjust the compression parameters accordingly.

The results demonstrate that the 2D U-Net model can perform brain tumor segmentation in MRI images but highlights areas for improvement. The discrepancy between training and validation metrics suggests overfitting, which could be addressed by incorporating more robust regularization techniques, data augmentation, or using more complex models such as 3D U-Nets. Additionally, further hyperparameter tuning and exploring different loss functions may enhance the model's performance.

Finally, a thorough evaluation of the proposed method in real-world telemedicine settings is crucial. This would involve assessing the impact of the compression on diagnostic accuracy and clinician satisfaction, as well as exploring the method's performance in live transmission scenarios where bandwidth and latency may vary. These steps will be critical in advancing the adoption of ROI-based compression in practical healthcare applications.